\documentclass[twocolumn,times]{aastex62}
\usepackage{amsmath}
\usepackage{amssymb}
\usepackage{lineno}
\allowdisplaybreaks[4]

\usepackage{color}

\shorttitle{Retrograde ring formed around eccentric extrasolar giant planet}
\shortauthors{Wenshuai Liu}
\begin{document}

%\title{Slow radial migration of a gap-opening planet triggered by dust feedback}
\title{\large{\textbf{Retrograde Ring Formed Around Eccentric Extrasolar Giant Planet}}}

\correspondingauthor{Wenshuai Liu}
\email{674602871@qq.com}

\author{Wenshuai Liu}
\affiliation{School of Physics, Henan Normal University, Xinxiang 453007, China}



\begin{abstract}
We investigate the accretion flow around a giant planet using two-dimensional hydrodynamical
simulations by studying the local region of accretion disk around the planet. The results show that, when the initial orbit of the planet embedded in protoplanetary disk is eccentric, the accretion disk formed around the
planet is retrograde during the evolution and may be a possible origin of the retrograde ring around eccentric extrasolar giant planet.
\end{abstract}



\section{Introduction}

Circumstellar disks consisting of gas and dust are the birthplace of planetary system. When the mass of the growing protoplanet reaches above about a few times as large as that of Earth \citep{1}, gas will accrete onto the protoplanet and a circumplanetary disk forms inside the Hill sphere where the protoplanet's gravity acting on the disk material is larger than that of the host star, leading to formation of gas giant planet and the redistribution of angular momentum and mass from the circumstellar disk material. Moons can form from the circumplanetary disk outside of the planet's Roche limit inside which ring system may exist \citep{2,3}. Observations show the existence of several tens of days of eclipses, meaning a circumplanetary ring system around the planetary companion may be possible \citep{4,5,6,7}. The direction of spin of the circumplanetary disk may be diverse in the extrasolar system. According to recent research \citep{8}, there may exist a retrograde ring system around J1407b. Thus, formation of retrograde ring or circumplanetary disk should be investigated. Here we investigate such situation from simulation of interaction of an eccentric massive planet and a gas disk using two-dimensional hydrodynamical simulations and further study the local region of accretion disk around the planet. Then we describe the flow pattern around the planet inside the planet's Hill sphere in detail. The results show that, when the eccentricity of the planet is sufficiently large, the accretion disk formed around the massive planet is retrograde during the evolution.

Planetary rings are common in solar system, with Saturn the most obvious one, being the evidence of circumplanetary disk during accretion history. In analogy with our own solar system, we may expect that similar structure should be around extrasolar planet. Detection of extrasolar planet usually analyzes the light curve by the transits of giant planet \citep{9,10}. If there exists ring system around such giant planet, the transits will produce distinctly detectable light curve with radius of ring several times that of host planet \citep{11,12}. It has been recently suggested that an extrasolar ring-like structure surrounding J1407b may be in the retrograde direction \citep{8}, raising question about its origin. According to many planet-disk hydrodynamical simulations, prograde circumplanetary disk usually forms around giant planet which rotates around its host star in prograde direction \citep{13}. What is it like in the situation of circumplanetary formed around eccentric planet interacting with circumstellar disk?

Doppler exoplanet surveys show the presence of many Jupiter-mass planets with orbit of large eccentricity \citep{14}, spanning from near-zero to near-unity. Planet-disk interaction tends to circularize the orbit of the newly formed planet embedded in protoplanetary disk, implying that giant planets formed through core accretion prefer to circular orbit. Furthermore, giant planets formed by gravitational instability have eccentricity up to 0.35 \citep{15}. In order to account for the observation, some post-formation dynamical mechanisms should be proposed to reproduce the eccentricity distribution of extrasolar giant planets. Dynamical instability caused by planet-planet scattering in multiplanet systems could excite eccentricity growth and may result in planet injection and merger \citep{16,17}. If the disk is still exist during such violent gravitational interaction within the planets, planet-disk interaction will go on. When one planet acquires extreme eccentricity relative to others after scattering, we may expect that the planet will be isolated from others due to the quick increase of its pericenter distance through interaction with disk, then, perturbations from other planets can be neglected \citep{18}. Some researches show that, after planet-planet scattering, there will be one giant planet left with large eccentricity up to 0.8 \citep{16}. If such planet is still embedded in disk, interaction between the eccentric planet and disk may form disk around the planet. Beside the mechanism of planet-planet scattering which drives up planetary eccentricity, Kozai-Lidov oscillations \citep{19,20,21}, stellar flyby \citep{22} and planet-disk interaction \citep{23,24} can also change eccentricity of planet. Studies in \cite{100} show that the eccentricity of giant planet with mass of several tens of the mass of Jupiter could reach 0.7 during planet-disk interaction.

Many extrasolar massive planet with mass of Jupiter are observed with large eccentricity \citep{14}. Usually, low mass planets found in extrasolar system have low eccentricity, indicating that they formed by accretion of gas and dust onto the embryo core embedded in an accretion disk. Thus, the final planet will have nearly circular orbit. Massive planet companion with brown dwarf mass couldn't form by core accretion due to the long timescale. Modern theories show that fragmentation of collapsing protostellar clouds caused by gravitational instability may be a viable way of the formation of brown dwarf mass planet \citep{25,26}, which may lead to the large eccentricity depending on the relative position and velocity of the fragments of the collapsing clouds. In contrast with the above, interaction of planet with circumstellar disk also drives the eccentricity of the planet. According to early research \citep{24}, the circumstellar disk could either increase or decrease the planet's eccentricity, which depends on the planet's mass relative to the host star. With respect to low mass planet, damp on the eccentricity usually occurs and the orbit is circularized during interaction with disk, while eccentricity are excited in the case of brown dwarf mass planet.

This work is organized as follows. We describe the model and initial conditions in Section 2. The results and discussions are given in Section 3.

\section{Initial Conditions}

Interaction of eccentric massive planet and circumstellar disk may be common in the extrasolar planetary system, especially in binary system containing a brown dwarf mass companion. Prograde circumplanetary disks usually form around giant planets embedded in disk around its host star. What would happen when turned to eccentric giant planet? To answer this question, we investigate the situation where an eccentric planet produced by some eccentricity exciting process, gravitationally interacts with the circumstellar disk in which it is embedded.

\begin{figure*}
   \begin{center}
     \begin{tabular}{cc}
            \includegraphics[width=0.5\textwidth]{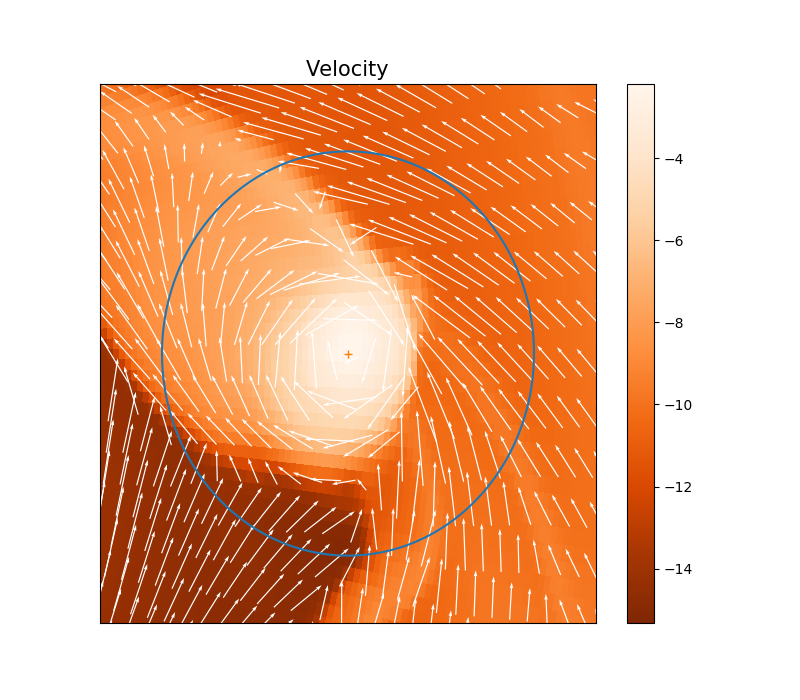}
            \includegraphics[width=0.5\textwidth]{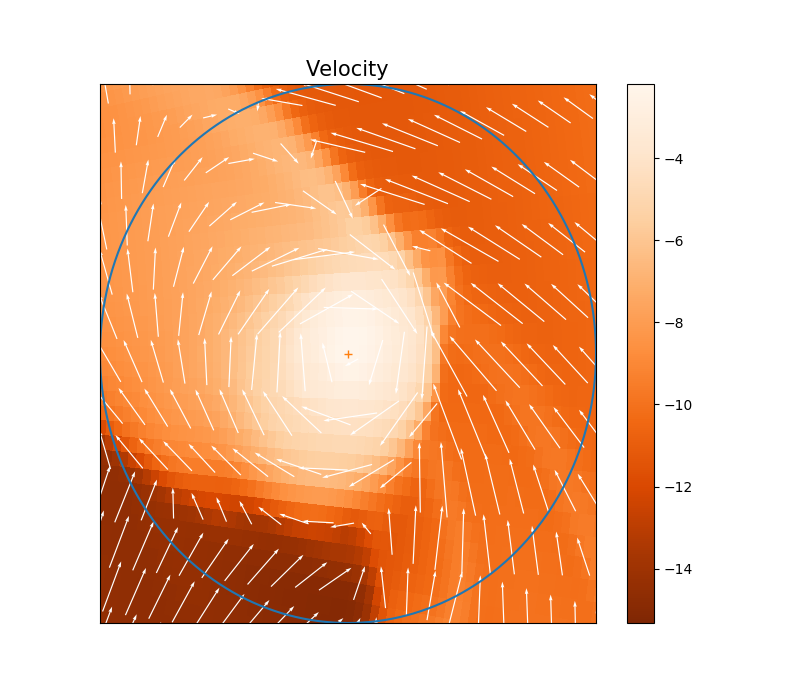}
            \end{tabular}
   \end{center}

\caption{ Left panel shows gas density according to the colorbar which is in logarithms and accretion flow around massive planet with $M_p=0.02$ in unit of mass of the sun, eccentricity $e=0.6$, and semimajor $a=1$ in unit of 5 AU. Right panel shows gas density according to the colorbar which is in logarithms and accretion flow around massive planet with $M_p=0.02$ in unit of mass of the sun, eccentricity $e=0.6$, and semimajor $a=1$ in unit of 30 AU. The blue circle is the Hill sphere of the massive planet. We can clearly see that retrograde circumplanetary disk forms around such planet.}
\label{fig:figure1}
\end{figure*}

The two-dimensional simulations are conducted using hydrodynamics with planet of mass $M_p=20 M_{Jupiter}$ embedded in disk around host star of mass $M_s=M_{sun}$. The planet, with eccentricity of 0.6, has semi-major of 5 AU and 30AU (AU is the astronomical unit) from the host star in the two simulations, respectively. We use the publicly available hydrodynamical code FARGO3D \citep{30} to simulate the two-dimensional planet-disk interaction on a polar mesh 500$\times$500 which spans in azimuth [-$\pi$,$\pi$] and in radius [0.1,5.0] in unit of the semimajor of the massive planet. In simulation One: the planet mass is $M_p=0.02$ in unit of mass of the sun, eccentricity $e=0.6$, and semimajor $a=1$ in unit of 5 AU. In simulation Two: the planet mass is $M_p=0.02$ in unit of mass of the sun, eccentricity $e=0.6$, and semimajor $a=1$ in unit of 30 AU. The surface density of the circumstellar disk is $\Sigma=3.1*10^{-5}(\frac{r}{a})^{-0.5}$ in code units and the aspect ratio is 0.05. We set the uniform kinematic viscosity of the gaseous disk to be $10^{-5}$.

\section{Results and Discussions}
Figure.~\ref{fig:figure1} shows the flow pattern around the giant planet after about 100 orbits to reach a quasi-steady state in each simulation. We can clearly see that flow in the inner of Hill sphere rotates around the planet in the retrograde direction. This simulation is conducted with planet's eccentricity of 0.6. Further simulations with eccentricity larger than 0.6 show similar flow pattern, indicating that gravitational interaction of massive planet with large eccentricity and circumstellar disk could result in retrograde circumplanetary disk. Simulations in this work are done with pure hydrodynamics without dust. When considering dust in the disk, the dusts will flow along the gas due to gas drag. After the dispersion of gas, a dusty ring is left around the massive giant planet and spreads inward and outward due to interparticle collisions, which may produce light curve if radius of the ring system is several times as large as that of its host planet and the angel between the plane of the ring and that of the orbit of the host planet is sufficiently large. Thus massive planets with sufficiently large eccentricity tend to have retrograde circumplanetary disk or ring. With a large amount of accretion of angular momentum from the retrograde circumplanetary disk, the final spin of the giant planet may also be in the retrograde direction. Retrograde moon may form in the retrograde ring and can survive with a longer time than that on prograde orbit due to the increased relative stability \citep{27,28}. The direct ditection of spin of circumplanetary disk or ring could test the result in this work and and provide a tool to determine the spin of the host planet since after large amount of accretion of disk material, the host planet spin may align with its ring. Simulations with much more massive planet of different semimajor radius and larger eccentricity are also done, with results showing that retrograde circumplanetary disk can form around its host planet.

Circumplanetary disks are expected to form around host planet with prograde spin with respect to the circumstellar disk. Results form simulations in this work prefer to retrograde circumplanetary disk around massive planet with large eccentricity and may be a possible origin of the retrograde ring around J1407b shown in recent research \citep{8}. The direct detection of spin axis of circumplanetary disk or ring around extrasolar massive planet will provide a complementary way to confirm results in this work. Theoretical works \citep{29} demonstrate that supermassive binary black holes with intermediate mass ratio and large eccentricity will form after merge of galaxies according to hierarchical model. Similar properties as in this work will exist if such supermassive binary black holes interact with the disk they are embedded in.

%\bibliographystyle{aasjournal}
%\bibliography{reference} % if your bibtex file is called example.bib

\end{document}